\documentclass[10pt,letterpaper]{article}
\usepackage{opex3}
\usepackage{epstopdf}

\usepackage{verbatim} 

\usepackage{amsmath,amsfonts,amssymb}
\usepackage{upgreek}

\begin{document}

\title{Optical Properties of Superconducting Nanowire Single-Photon Detectors}

\author{Vikas Anant,$^1$ Andrew J.\ Kerman,$^2$ Eric A.\ Dauler,$^{1,2}$ Joel K.\ W.\ Yang,$^1$ Kristine M.\ Rosfjord,$^3$ and Karl K.\ Berggren$^1$}
\address{$^1$Research Laboratory of Electronics, Massachusetts Institute of Technology, Cambridge, Massachusetts 02139, USA\\
$^2$Lincoln Laboratory, Massachusetts Institute of Technology, Lexington, Massachusetts 02420, USA\\
$^3$Institute for Research in Electronics and Applied Physics, University of Maryland, College Park, Maryland 20742, USA}
\email{berggren@mit.edu}

\begin{abstract}
We measured the optical absorptance of superconducting nanowire single photon detectors.  We found that 200-nm-pitch, 50\%-fill-factor devices had an average absorptance of 21\% for normally-incident front-illumination of 1.55-$\upmu$m-wavelength light polarized parallel to the nanowires, and only 10\% for perpendicularly-polarized light.  We also measured devices with lower fill-factors and narrower wires that were five times more sensitive to parallel-polarized photons than perpendicular-polarized photons.  We developed a numerical model that predicts the absorptance of our structures.  We also used our measurements, coupled with measurements of device detection efficiencies, to determine the probability of photon detection after an absorption event. We found that, remarkably, absorbed parallel-polarized photons were more likely to result in detection events than perpendicular-polarized photons, and we present a hypothesis that qualitatively explains this result. Finally, we also determined the enhancement of device detection efficiency and absorptance due to the inclusion of an integrated optical cavity over a range of wavelengths (700-1700 nm) on a number of devices, and found good agreement with our numerical model.
\end{abstract}


\section{Introduction}

The short reset-time, low jitter, and broad wavelength response of superconducting nanowire single photon detectors (SNSPDs) makes them attractive candidates to replace other single photon detectors, including avalanche photodiodes \cite{apd}, in applications such as free-space optical communications \cite{eric}, quantum cryptography networks \cite{quantum-crypto,takesue}, and quantum computation \cite{knill}.  To fully take advantage of the boost in speed afforded by these properties, SNSPDs with high system detection efficiencies ($SDE$) are needed.  But SNSPDs typically exhibit $SDE$s of only 0.2-10\% for 1.55-$\upmu$m-wavelength single-photons \cite{verevkin,korneev,hadfield}.  This limitation is currently believed to be due to poor coupling efficiency ($\eta_\text{c}$); however, this may not be the entire story.  To understand where the losses come from, we need to examine the factors that contribute to the $SDE$.

The $SDE$ is a product of two lumped quantities, the coupling efficiency ($\eta_\text{c}$) and the device detection efficiency ($DDE$).  The quantity $\eta_\text{c}$ encapsulates all loss mechanisms encountered between the photon source and SNSPD, and is defined as the ratio of the number of photons that reach the active area to the number of photons emitted by the photon source.   It is relatively straightforward to realize $\eta_\text{c}>0.9$ through careful optical design that minimizes these losses.  On the other hand, maximizing $DDE$ is not as simple.

The $DDE$ is the probability that a photon incident on the active area results in a voltage pulse.  $DDE$ depends on two quantities, the absorptance $A$, and the probability of resistive state formation due to an absorption event $P_\text{R}$ through 
\begin{equation}
  DDE=P_\text{R}\, A.
  \label{eq.DE}
\end{equation}
While $P_\text{R}$ depends on the microscopic physics of the nanowires, $A$ depends only on the optical properties of our fabricated structures and the incident field.  $A$ needs to be maximized to make efficient detectors; however, there are many ways an incident photon can remain unabsorbed.  For example, the photon can pass through open gaps between the nanowires or be transmitted through the subwavelength thickness of NbN.  In addition, a low value for $P_\text{R}$ will further suppress $DDE$ (and therefore also $SDE$).

In this paper, we experimentally demonstrate that the optical polarization and changes in the geometry (pitch and fill-factor) of SNSPDs impact their absorptance and efficiency.  For example, for devices with 200 nm pitch and 50\% fill factor, we found that 21\% of light incident from the front was absorbed for parallel polarization, while only 10\% was absorbed for perpendicular polarization.  The absorptance was reduced to 14\% and 6\% for parallel and perpendicular polarizations, respectively, for devices with the same pitch but a 25\% fill-factor.  We describe both a numerical model that predicts the absorptance of our structures, and experiments that directly measure absorptance.  We found that the experimental data matches the predictions of our model to within the uncertainties in our knowledge of physical and optical parameters.  We also measured the $DDE$ for the same devices and compared it to the measured absorptance.  We discovered, remarkably, that $P_\text{R}$ is different for photons polarized perpendicular ($\perp$) to the nanowires ($P_{\text{R},\perp}$) than for photons polarized parallel to the nanowires ($P_{\text{R},||}$).  We propose a model that qualitatively explains our observation of $P_{\text{R},||} \neq P_{\text{R},\perp}$.

Finally, we also made measurements to confirm that our numerical model and modeling parameters are accurate over a range of wavelengths.  We measured the factor of enhancement of $DDE$ due to the addition of an integrated optical cavity for single-photon wavelengths of 700-1700 nm, and found good agreement with our numerical model.  As a result, we have shown that optical modelling can be used to design future SNSPDs in new regimes of operation.

This paper is organized as follows.  In Section \ref{sec.model}, we will describe a numerical model that predicts the absorptance of our structures.  In Section \ref{sec.expt}, we will describe the setup and experiments conducted to measure absorptance on our devices.  In Section \ref{sec.wavelength}, we will describe the experiments conducted to measure the wavelength dependence of the enhancement due to the addition of an optical cavity.  We will not describe fabrication details or $DDE$ measurement details as these were discussed in \cite{joel} and \cite{rosfjord}, respectively.

\section{Optical Model} \label{sec.model}

We modeled the absorption process of photons by an SNSPD as a plane wave interacting with an infinite grating.  These two approximations are justified because in our experiments, the photon was launched at normal incidence and had plane-wave phase-fronts when it reached the detectors, and because our beam spot was smaller than the test grating which will be described in Section \ref{sec.expt}. 

While analytic methods, such as form birefringence theory, have been used in the past to analyze grating structures such as wire-grid polarizers \cite{wgp-yu,wgp-yeh}, they break down for subwavelength grating thicknesses, in particular for electric-field polarized perpendicular to the wires in the grating (i.e. in the direction of grating periodicity)\footnote{For an electric field polarized parallel to the wires in a subwavelength grating, an accurate result for the $A_{||}$ can be obtained with the Fresnel equations where an effective index $n_\text{eff} = ((1-f)\,n_\text{air}^2 + f\,n_\text{NbN}^2)^{1/2}$ is used for the thin film consisting of NbN subwavelength gratings ($n_\text{air}=1$, $n_\text{NbN}=5.23-i\,5.82$, and $f$ is the fill-factor).  A simple effective index model that only depends on $f$ will not work for perpendicular polarization since $A_\perp$ depends on both $f$ and $p$.} \cite{lalanne}.  Recently, a numerical technique (finite-element analysis) was used to approach this problem \cite{majedi}; but because of either a difference in material parameters or simulation conditions, their findings were not consistent with our experimental or theoretical results.

\begin{figure}[htbp]
  \centering
  \includegraphics[angle=90]{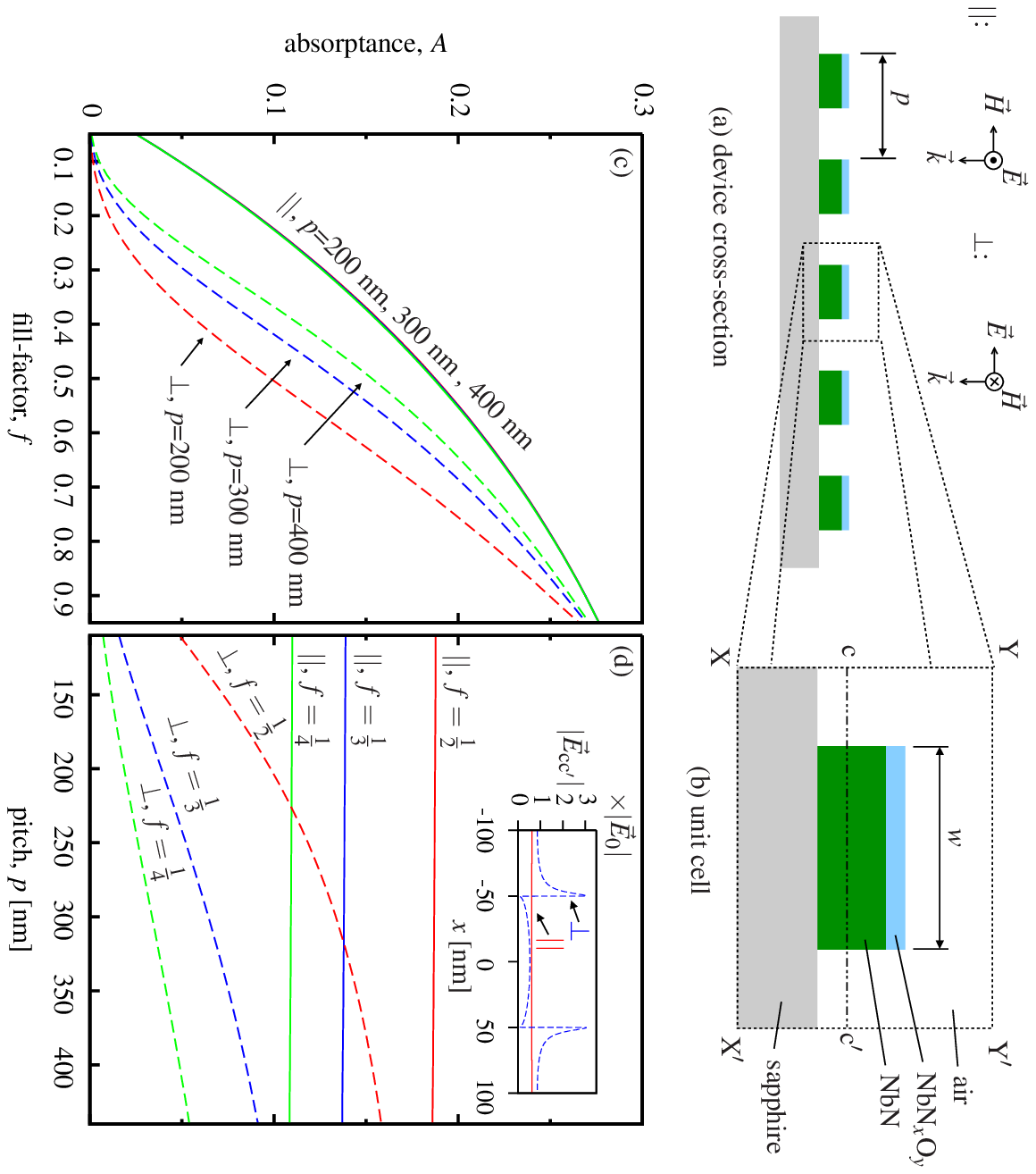}
  \caption{Infinite wire grid schematized in (a) is further reduced via symmetry to the unit cell shown in (b) in numerical modeling of the absorptance, $A$.  (The schematic is not drawn to scale.)  Plots of the calculated $A$ as a function of fill-factor, $f$ and pitch $p$ are shown in (c) and (d) for parallel ($||$) and perpendicular ($\perp$) electric field polarization.  Inset in (d) shows how the calculated cross-sectional time-averaged electric field magnitude, $|\vec{E}_{\text{cc}^\prime}|$ varies with position $x$ for $||$ and $\perp$ polarizations across a $f=\frac{1}{2}$, $p=200\,\text{nm}$ structure.  The NbN film thickness ($t_\text{NbN}$) was 4 nm and incident electric field magnitude $|\vec{E}_\text{0}|=1$.  The NbN region extends from $x=-50\,\text{nm}$ to $x=+50\,\text{nm}$.}
  \label{fig.snspd-geometry}
\end{figure}

We used a different finite-element analysis software (Comsol Multiphysics v3.2b, EM module) than the one used in \cite{majedi} to calculate the absorptance of the geometry shown in Fig. \ref{fig.snspd-geometry}(a).  We took advantage of symmetry by using the `In-plane Hybrid-Mode Waves' implementation and defined the unit cell\footnote{While our fabrication process \cite{joel,rosfjord} leaves behind 10-40 nm of residual resist on top of the nanowires, we found that including the resist in our geometry did not affect our results.} shown in Fig. \ref{fig.snspd-geometry}(b) with the desired NbN thickness ($t_\text{NbN}$), pitch ($p$), wire width ($w$), fill-factor ($f=w/p$), and a 1.55-$\upmu$m-wavelength.  The thickness of NbN$_x$O$_y$ was held constant at \mbox{2 nm} because it was observed by transmission electron microscopy (TEM) to be the same thickness for different NbN thicknesses.\footnote{TEM imaging services were provided by Materials Analytical Services, Inc.}  The sapphire wafer thickness was not included because an anti-reflection coating was added to the sapphire-air interface in experiments.  We applied periodic boundary conditions on the  left (XY) and right ($\text{X}^\prime\text{Y}^\prime$) edges of the unit cell, scattering boundary conditions on the bottom (XX$^\prime$) and top (YY$^\prime$) edges.  We applied either an out-of-plane ($E_{||}$), or in-plane ($E_\perp$) electric field on the top edge (front-illumination), which matches the experimental setup described in Sec.\ \ref{sec.expt}.  The material parameters were defined in terms of complex refractive indices collected either from measurements\footnote{Measurements of the refractive indices of NbN and $\text{NbN}_x\text{O}_y$ made at room temperature by J.\ A.\ Woolam, Inc.\ using spectroscopic ellipsometry on a 12-nm-thick \uppercase{N}b\uppercase{N} film deposited on a sapphire wafer.} or the literature.  We used \mbox{$n_\text{NbN}=5.23-i\,5.82$,} \mbox{$n_{\text{NbN}_x\text{O}_y} = 2.28$,} and \mbox{$n_\text{sapphire} = 1.75$ \cite{malitson}.} We then applied a Lagrange (quadratic) mesh constrained to 0.25 nm in NbN and $\text{NbN}_x\text{O}_y$, and made to be denser at edge XY than $\text{X}^\prime\text{Y}^\prime$ to maintain accuracy when using periodic boundary conditions.  Moving to a denser mesh did not improve accuracy.  We used a direct linear solver (UMFPACK) to solve for the spatial distribution of electric ($\vec{E}$) field, $\vec{E}(x,y)$.

The absorptance $A$ is proportional to the integral of the time-averaged electric field intensity in the NbN film.  Because the electric field does not vary significantly over the thickness of the NbN, we can express $A$ in terms of the cross-sectional electric field $|\vec{E}_{\text{cc}^\prime}(x)|$ in NbN $\big($as pictured in Fig.\ \ref{fig.snspd-geometry}(b)$\big)$ through
\begin{equation}
  A = \int_{-p/2}^{p/2}\,\int_0^{t_\text{NbN}} Q(x,y)\, \text{d}x\, \text{d}y\, \bigg/ \int_{-p/2}^{p/2}\, I_\circ\, \text{d}x
\end{equation}
where $Q(x,y)$ is the time-average resistive dissipation in the nanowire, and $I_\circ$ is the time-average incident Poynting power density given by the following equation
\begin{equation}
  I_\circ = \frac{1}{2}\,(\epsilon_\circ/\mu_\circ)^{1/2}\, |\vec{E}_\circ|^2
\end{equation}
where $|\vec{E}_\circ|$ is the time-averaged incident electric field magnitude, $\epsilon_\circ$ is the permittivity of air, and $\mu_\circ$ is the permeability of air.   For our wires which have typical $t_\text{NbN}$ of 4-6 nm, $Q(x,y)\approx Q(x)$ and is given by Ohm's law to be
\begin{equation}
  Q(x) = \frac{1}{2}\, \omega\, \text{Im}[\epsilon] |\vec{E_{\text{cc}^\prime}}(x)|^2.
\end{equation}

Figure \ref{fig.snspd-geometry}(c) shows the calculated dependence of $A$ on $f$ for constant values of $p$ and electric field polarizations that were either parallel ($||$) or perpendicular ($\perp$) to the periodicity of the nanowires, while Fig.\ \ref{fig.snspd-geometry}(d) shows $A$ as a function of $p$.  Both figures show that $A_{||}$ is invariant with $p$ when $f$ is kept constant, and that $A$ can be maximized by narrowing the gaps between nanowires.  We understand that $A_{||}$ remains constant with varying $p$ because the electric field intensity is continuous across the air-NbN boundaries as a result of tangential $\vec{E}_{\text{cc}^\prime}^{||}$ continuity, as shown in Fig. \ref{fig.snspd-geometry}(d) inset.  Only the fraction of the electric field magnitude $|\vec{E}_{\text{cc}^\prime}^{||}|$ in NbN compared to air (i.e. $f$) that changes $A_{||}$.  On the other hand, decreasing $p$ while keeping $f$ constant decreases $A_\perp$.  This effect is observed because there are more air-NbN interfaces per unit area for a grating with a smaller pitch than one with a larger pitch, and the boundary conditions dictate a lower $|\vec{E}_{\text{cc}^\prime}^{\perp}|$ in the NbN at each air-NbN interface.

The plots also illustrate an important limitation that $A$ poses to the photon detection process: the maximum $A$ (and therefore also $DDE$) can not exceed 30\% with the configuration shown.  There are two ways that $A$ can be increased that have already been demonstrated: (1) by using back-illumination (i.e. through sapphire edge $\text{XX}^\prime$) instead of front-illumination (i.e. through air edge $\text{YY}^\prime$), and thereby reducing the index mismatch with NbN, $A$ can be increased up to 45\%; (2) by fabricating an optical cavity designed to intensify the field in the NbN nanowires \cite{rosfjord}.  An understanding of the impact of SNSPD geometry and optical polarization on $A$ and $P_\text{R}$ will be helpful in finding other ways to improve $A$. 

We will now describe how we determined the optical absorptance of the devices and how the results compared to the model.  We will also compare our measurements of the absorptance to the $DDE$ measurements of the same devices.

\section{Absorptance measurements}\label{sec.expt}

\begin{figure}[htbp]
  \centering
  \includegraphics[angle=90]{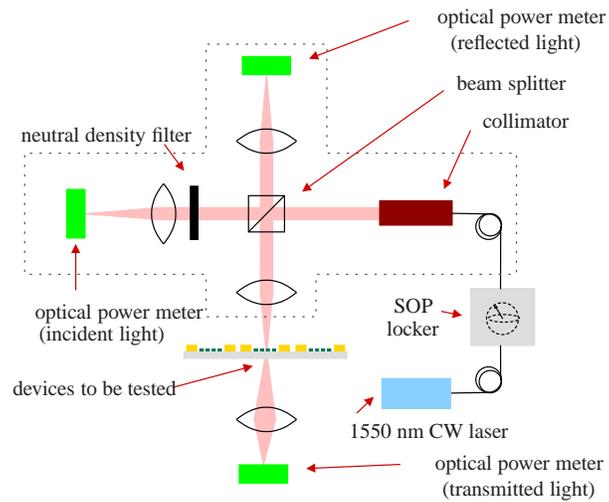}
  \caption{Schematic of the optical setup used to measure the optical absorptance.  A state of polarization (SOP) locker was used to set the input polarization of the incident 1.55-$\upmu$m-wavelength light.  Light was focused to a 4 $\upmu$m (FWHM) spot onto a device and the incident, reflected, and transmitted power were measured.  The absorptance was calculated from these measurements and earlier calibrations discussed in the main text.}
  \label{fig.snspd-testing-schematic}
\end{figure}

Figure \ref{fig.snspd-testing-schematic} shows a schematic of the optical apparatus we used to measure absorptance.  Free-space optics were attached to a three-axis, computer-controlled stage (motorized actuator: Newport LTA-HS, motion controller: Newport ESP300).  The stage could be affixed to a cryostat specially designed to allow free-space optical coupling to a cold sample, or be used without a cryostat for room-temperature measurements.  We mounted a single-mode optical fiber (Corning SMF-28) to the stage that was illuminated by a 1.55-$\upmu$m-wavelength diode laser (Thorlabs S1FC1550) and polarized using a state of polarization (SOP) locker (Thorlabs PL100S).  The SOP locker was preferable to a manual polarizer because it provided mechanical isolation to our setup when we varied polarization and permitted automation of the data acquisition process. The polarized light was collimated and split equally into two arms.  In one arm, a Ge-based sensor (Thorlabs S122A) connected to a power meter (Thorlabs PM100) was used to measure the incident light intensity, mainly for troubleshooting purposes.  In the other arm, a long-working-distance microscope objective (Mitutoyo M Plan APO NIR 20X) was used to focus the light onto a device with a spot-size of $4\, \upmu\text{m}$ full-width half-max (FWHM).  Light reflected from the device was collected and measured using another power meter.  The transmitted light was measured by a third power meter.  We controlled the power meters, SOP locker, and motion controller using the Instrument Control Toolbox (ver.\ 2.4) for Matlab (ver.\ 7.2/R2006a) and custom software. 

\begin{figure}[htbp]
  \centering
  \includegraphics[angle=90]{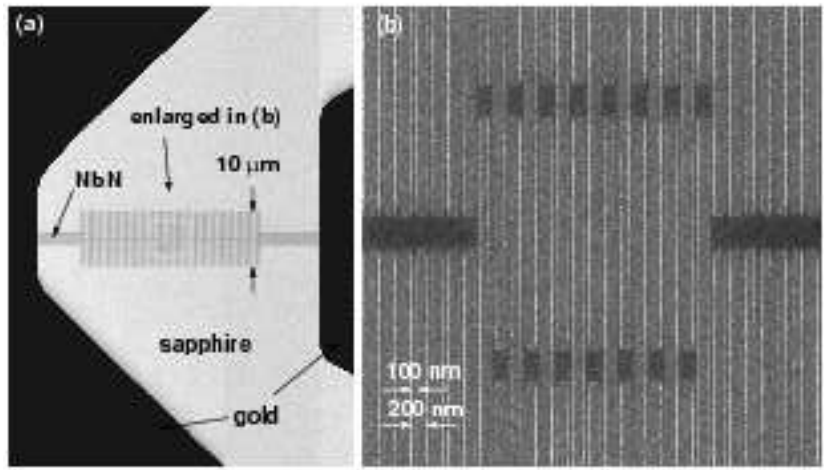}
  \caption{Scanning electron micrograph of a 50\% fill-factor, 200-nm-pitch SNSPD is shown in (a).  The central, photon-detecting region of the SNSPD structure outlined in (a) is magnified in (b).  The extended parallel grating structure outside the active area was necessary for proximity-effect correction in the fabrication process, but was also useful in expanding the optically testable region of the device.}
  \label{fig.sems}
\end{figure}

It was necessary to fabricate devices which were specially designed to facilitate both absorptance and $DDE$ measurements.  A typical device is pictured in Figure \ref{fig.sems} where the active area ($3\, \upmu\text{m} \times 3\, \upmu\text{m}$) of the SNSPD is centrally located within a large grating structure ($30\, \upmu\text{m} \times 10\,\upmu\text{m}$).  The small active area facilitated $DDE$ measurements because it enabled better uniformity in electrical response across the device.  The larger grating structure, which has the same $p$ and $f$ as the smaller active area, facilitated measurements of $A$ because it closely matched our simulation geometry.\footnote{98\% of the laser intensity in a 4 $\upmu$m FWHM laser spot was incident on the large grating structure.}  Thus, five groups of devices with identically sized large gratings and small active areas were fabricated with different $(f,\, p\, \text{[nm]})$ combinations of $(\frac{1}{2}, 200)$, $(\frac{1}{3}, 300)$, $(\frac{1}{4}, 400)$, $(\frac{1}{3}, 150)$, and $(\frac{1}{4}, 200)$ on the same chip.

The procedure for collecting absorptance data was as follows.  We first measured the reflected and transmitted power from gold and sapphire in the vicinity of the device.  These measurements were repeated over polarizations spaced by 10$^\circ$ on the Poincar\'{e} sphere.  We calibrated the measured reflected and transmitted power to the theoretical reflectance and transmittance $(R,\, T)$ of gold\footnote{The patterned gold film was approximately 100 nm thick.  We used bulk values for the complex refractive index of gold found in \cite{palik}.  For a thin film thicker than 30 nm, the bulk refractive index can be used in this wavelength range \cite{reale}.} $(0.98,\, 0)$ and sapphire $(0.07,\, 0.93)$ calculated from their refractive index values at $\lambda=1.55\,\upmu\text{m}$.  We then measured the reflected and transmitted power with the laser centered in the active area of the SNSPD, and calculated $R$, $T$, and $A=1-R-T$.  We then repeated our measurements on 10-13 identical devices in each of the five groups of devices. 

\begin{figure}[htbp]
  \centering
  \includegraphics[angle=90]{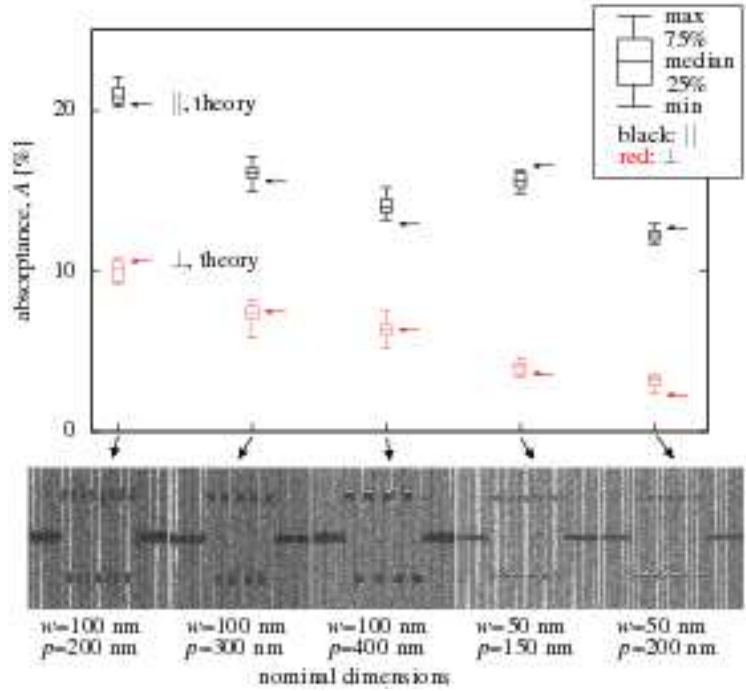}
  \caption{Statistical plot of the measured parallel ($||$) and perpendicular ($\perp$) absorptance for devices with different pitch $p$ and fill-factor $f$.  Each data symbol represents measurements of 10-13 devices for the polarization that yielded the maximum ($||$) or minimum ($\perp$) absorptance.  The arrows indicate the calculated absorptance for structures with fitted wire widths (from left to right) $w$=104 nm, 108 nm, 114 nm, 58 nm, and 55 nm, NbN thickness \mbox{$t_\text{NbN}$=4.35 nm}, and nominal values of pitch $p$.}
  \label{fig.A-vs-device-type}
\end{figure}

Figure \ref{fig.A-vs-device-type} shows our measurements of $A_{||}$ and $A_\perp$ at room temperature.  Room temperature measurements did not differ significantly from measurements of $A$ made at 6 K.  We attempted to fit the model by using $w$ and $t_\text{NbN}$ as free parameters, and assuming $t_\text{NbN}$ was between \mbox{4 nm} and \mbox{6 nm}.  Our fit required wire widths that were systematically 10-15\% larger than the nominal $w$ used in our electron-beam-lithography pattern.  There are many uncertainties in our model parameters that can contribute to this.  One cause is that we may have inadvertently exposed the resist adjacent to the intended structure causing $w$ to be larger than desired.  We conducted scanning electron microscopy of our nanowires and found that $w$ predicted by the model were within the uncertainty caused by increased secondary electron emission from the edges of the nanowires.  However, there are other factors that can contribute to a fitting error.  One factor is that the NbN refractive index we used in our simulations may not be accurate because the refractive index was measured on a thicker film (12 nm) than the film we used to fabricated our devices on (4-6 nm).  In addition, the films were from different growth batches so may have had slight differences in optical properties.  Another factor is that our simple model of infinite gratings may not be sufficient, i.e.\ a 3D model may be needed.  Considering these uncertainties, the qualitative and quantitative fit of our measurements with the model was good. 

\subsection{Comparison of device detection efficiency to absorptance}

\begin{figure}[htbp]
  \centering
  \includegraphics[angle=90]{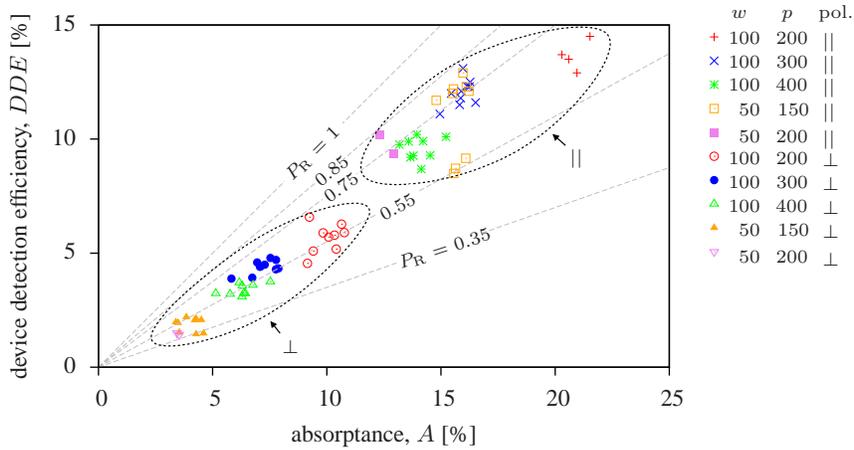}
  \caption{Plot of device detection efficiency as a function of absorptance for the same devices.  Dotted lines have a constant slope given by $P_\text{R}$, the probability of resistive state formation.}
  \label{fig.snspd-testing-DE-vs-A}
\end{figure}

A comparison of device detection efficiency to absorptance can answer one important question about our device: is every absorbed photon actually detected?  To answer this question, we measured both $DDE$ (using the measurement apparatus described in \cite{rosfjord} but with front-illumination instead of back-illumination) and $A$ for the same devices.  The two quantities are plotted against each other in Fig.\ \ref{fig.snspd-testing-DE-vs-A} where, for reference, we have also plotted lines with constant slope $P_\text{R}$. It should be noted that the value for $DDE$ is measured with a relative accuracy of $\pm 25\%$ due to uncertainties in fiber output power calibration. Some features of this plot were as expected; for example, all of the devices had higher absorptance than device detection efficiency ($P_\text{R}<1$).  But there were other features that were unexpected.  We will now discuss these other features and see how they yield an unexpected new insight into the microscopic physics of SNSPDs.

There are two remarkable features in the data plotted in Fig.\ \ref{fig.snspd-testing-DE-vs-A}.  First, $P_\text{R}$ had a peak value of 0.85$\genfrac{}{}{0pt}{3}{+0.15}{-0.21}$ for parallel polarization.  The natural question one can ask is what happens to absorbed photons that do not lead to detection events?  One possible explanation is that the devices were in some way not biased at their "true" critical currents, i.e. that they were constricted in some way so that the absorption of photons does not cause a voltage pulse.  Another explanation is that there is some other absorbing medium that was not accounted for.  We aim to address this issue in future experiments.

A second remarkable feature of the data in Fig.\ \ref{fig.snspd-testing-DE-vs-A} is that $P_\text{R}$ for parallel polarization ($0.55\genfrac{}{}{0pt}{3}{+0.14}{-0.14} < P_{\text{R},||} < 0.85\genfrac{}{}{0pt}{3}{+0.15}{-0.21}$) is different than $P_\text{R}$ for perpendicular polarization ($ 0.35\genfrac{}{}{0pt}{3}{+0.09}{-0.09} < P_{\text{R},\perp} < 0.75\genfrac{}{}{0pt}{3}{+0.19}{-0.19}$).  While we do not yet completely understand the origin of the disparity between $P_{\text{R},||}$ and $P_{\text{R},\perp}$, we have developed a model that can form a starting point for further analysis.

The main assumption of this model is a position-dependent (but polarization-independent) probability \emph{density} of resistive state formation, $\psi(x)$, where $x$ is the distance across the nanowire.  The second element of this model is the time-averaged resistive dissipation $Q(x)$ which we know to be both position-dependent and polarization-dependent.  These two elements can be combined to give the device detection efficiency and absorptance for parallel and perpendicular polarization:
\begin{equation} \label{eq.DE-psi}
  DDE = \int_{-p/2}^{p/2}\int_0^{t_\text{NbN}} \psi(x) \frac{Q(x)}{I_\circ}\, \text{d}x\, \text{d}y.
\end{equation}
We can use this equation, Eq.\ \eqref{eq.DE}, and a calculated $Q(x)$ to find a $\psi(x)$ that gives a different $P_\text{R}$ for parallel and perpendicular polarizations.  Using our data, we found that $\psi(x)$ needs to be larger at the edges of the nanowire than in the center to explain $P_{\text{R},||}>P_{\text{R},\perp}$, however the five geometries we investigated did not provide enough resolution to fit $\psi(x)$ to a specific shape at this point.

\section{Verification of the model through measurement of the cavity enhancement factor}\label{sec.wavelength}

In the previous section, we verified that our model predicted the absorptance for our structures to within experimental uncertainties.  But this verification was done only at a single wavelength of 1550 nm, mainly because all of our optical components, i.e., the laser, fiber, lenses, SOP locker, microscope objective, beam splitter, collimator, and photodiodes were optimized for that wavelength.  In order to verify our modeling parameters for a range of wavelengths, we measured the cavity enhancement factor $\mathcal{E}$ for SNSPDs over a range of wavelengths.  The cavity enhancement factor $\mathcal{E}$ is the ratio of the absorptance of an SNSPD with an integrated cavity to the absorptance without a cavity and depends on the refractive indices of the materials and the geometry.  $\mathcal{E}$ is also the factor by which the intensity of the electric field in NbN, and therefore $Q(x)$, is increased due to the presence of a cavity.  In view of Eq.\ \eqref{eq.DE-psi}, and because $\psi(x)$ should not depend on the intensity, we can determine $\mathcal{E}$ by measuring $DDE$ for an SNSPD with a cavity and the $DDE$ of an SNSPD without a cavity and taking the ratio of these two quantities for each wavelength.

We generated radiation spanning the range from 600-1700 nm using a supercontinuum source (Toptica photonics). This source consisted of a power-amplified, modelocked fiber laser with 100 fs pulses at 1550 nm, which could either be coupled into a highly-nonlinear step-index fiber (generating a frequency comb from $\sim$600-1050 nm) or first doubled using a periodically-poled Lithium-niobate crystal and then coupled into a photonic crystal fiber (Blazed photonics) (generating a frequency comb from $\sim$ 1150-1700 nm). We used a  Pellin-Broca prism to select a wavelength band out of one or the other of these outputs (with a FWHM $\Delta \lambda/\lambda \sim 50$) which we then coupled into an optical fiber (SMF-28, single mode from $\sim$1200-1600 nm). The wavelength of the light output from the fiber could then be tuned by rotating the prism.

We used this tunable output to measure the enhancement $\mathcal{E}$.  The most obvious way to do this would simply be to make calibrated $DDE$ measurements at all wavelengths before and after adding the cavity. However, we chose not to attempt this, since we had no way to accurately calibrate optical power levels at or near the single-photon level over such a wide wavelength range. To avoid the need for this calibration, we instead measured only detection efficiency ratios for pairs of detectors side-by-side. To do this, we fabricated a chip with 225 pairs of 3$\times$3.3 $\upmu$m detectors, where the two in each pair could be read out separately, and were spaced by only 100 $\upmu$m.  We measured critical currents and room-temperature resistances of all 450 detectors, and identified $\sim$ 50 pairs for which these values were within 5\% of each other.  We then measured (calibrated) $DDE$s at 1550 nm for this subset of detectors, and further narrowed the experimental sample to the pairs having $DDE$s within 5\% of each other.  Then, we added optical cavities to one detector of each pair, and re-measured at 1550 nm.  Although we commonly observe that $DDE$ for a given device can vary between cooldowns (particularly if some additional processing has occurred), these variations are nearly always strongly correlated with a variation of the critical current (a reduced $DDE$ correlates with a reduced critical current).  So, we selected from our pairs of detectors only those where the critical currents remained the same to within 5\% after the addition of cavities, and where the detector which did not have a cavity added had the same $DDE$ as previously to within 5\%.  Lastly, we further reduced our experimental sample by selecting only the subset of devices which were relatively unconstricted \cite{jamie}.  In the present context, this corresponded to those devices with $DDE$ at 1550 nm $>$17\% for front illumination.  Our final sample consisted of five detector pairs.  For these ten detectors, we measured count rates at each of a sequence of wavelengths spanning the entire accessible range.  By measuring both detectors in a pair in succession without adjusting the source in any way, we ensured that the optical power incident on the two detectors was identical, and therefore that their count rates could be compared directly to give $\mathcal{E}$.  This precaution was important since the optical power obtainable in a given wavelength band was not repeatable, nor was the coupling into the SMF-28 optical fiber, which was not single-mode over a fraction of our wavelength range.  Given the latter issue, we restricted ourselves to the wavelength range 700-1700 nm where the fiber output could be fairly well polarized (extinction ratio $>$ 20). 

\begin{figure}[htbp]
  \centering
  \includegraphics[angle=90]{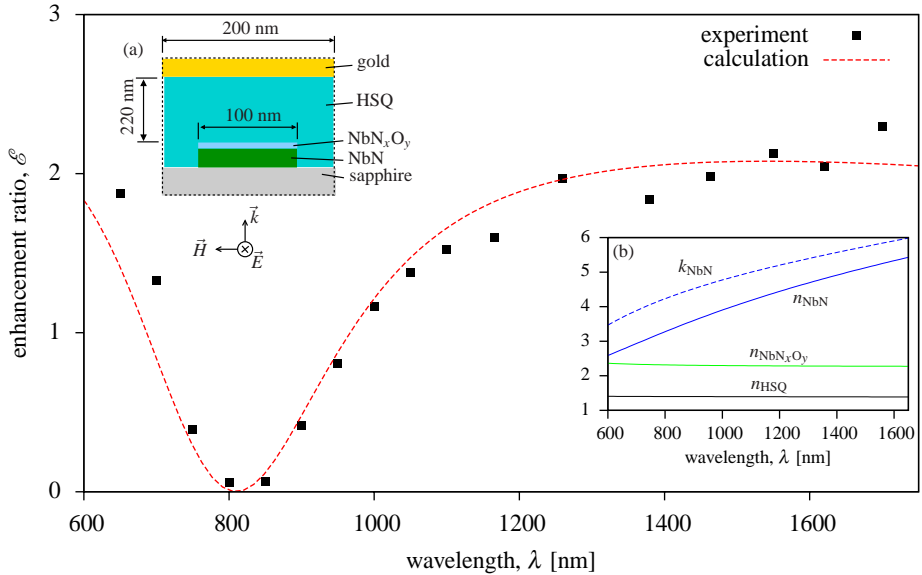}
  \caption{Plot of enhancement factor, $\mathcal{E}$ as a function of wavelength $\lambda$.  The dotted line shows the calculation that was carried out on the unit cell shown in inset (a) using values for the refractive indices shown in inset (b).}
  \label{fig.snspd-testing-DE-vs-wavelength}
\end{figure}

Figure \ref{fig.snspd-testing-DE-vs-wavelength} shows the enhancement measured in this manner, and a calculation of $\mathcal{E}$ as a function of wavelength.  We can see that the data has good experimental agreement with the calculation.  Inset Fig. \ref{fig.snspd-testing-DE-vs-wavelength}(a) shows the geometry that was used in the calculation while Fig. \ref{fig.snspd-testing-DE-vs-wavelength}(b) shows a plot of the real and imaginary parts of refractive index for NbN, NbN$_x$O$_y$, and HSQ\footnote{Measurements of the refractive indices of NbN and $\text{NbN}_x\text{O}_y$ were made at room temperature by J.\ A.\ Woolam, Inc.\ using spectroscopic ellipsometry on a 12-nm-thick \uppercase{N}b\uppercase{N} film deposited on a sapphire wafer.  Measurements of refractive index of HSQ were also performed by J.\ A.\ Woolam, Inc.\ using spectroscopic ellipsometry.  $n$ and $k$ for gold were taken from Ref.\ \cite{palik}}.

\section{Conclusion}

In this paper, we reported the first measurements of the absorptance of SNSPDs and showed how $A$ changed with the fill-factor and pitch of SNSPDs and optical polarization.  We confirmed that our numerical model made accurate predictions for a range of geometry, polarization, and wavelength.  We found that 200-nm-pitch, 50\% fill-factor devices had an average absorptance of 21\% for light polarized parallel to the nanowires, and only 10\% for perpendicularly-polarized light.  This disparity in polarization-sensitive absorptivity was even more evident in lower fill-factor and narrow wire-width devices, where we measured that parallel-polarized photons were more than 5 times as likely to be absorbed over perpendicularly polarized photons.  We also found that potentially, some absorbed photons do not result in detection events, and that this quantity is smaller for photons with an orthogonal polarization.  These results present new challenges that need to be understood and overcome if higher efficiency devices will be possible.

\section*{Acknowledgements}

The authors would like to thank Prof.\ Hank Smith for the use of his facilities and equipment, James Daley and Mark Mondol for technical assistance, and Ayman Abouraddy, Scott Hamilton, Rich Molnar, Prof.\ Steven Johnson, and Prof.\ Terry Orlando for helpful discussions. The authors would also like to thank Jeffrey Stern for useful discussions about using effective index theory for optical modeling and for pointing out the formula for $n_\text{eff}$ for parallel electric-field polarization. This work made use of MIT's shared scanning-electron-beam-lithography facility in the Research Laboratory of Electronics (SEBL at RLE).

This work was sponsored in part by the United States Air Force under Air Force Contract \#FA8721-05-C-0002 and IARPA.  Opinions, interpretations, recommendations and conclusions are those of the authors
and are not necessarily endorsed by the United States Government.

\bibliographystyle{osajnl}
\end{document}